\documentclass[preprints,article,accept,pdftex,moreauthors]{Definitions/mdpi} 

\usepackage{amssymb}
\usepackage{mathabx}
\usepackage{amsmath}

\newcommand{\beq}{\begin{equation}}
\newcommand{\eeq}{\end{equation}}
\newcommand{\bea}{\begin{eqnarray}}
\newcommand{\eea}{\end{eqnarray}}

\firstpage{1} 
\pubvolume{1}
\issuenum{1}
\articlenumber{0}
\pubyear{2023}
\copyrightyear{2022}
\externaleditor{}
\datereceived{} 
\daterevised{} 
\dateaccepted{} 
\datepublished{} 
\hreflink{https://doi.org/} 
\pdfoutput=1

\Title{Multiverse Predictions for Habitability: Stellar and Atmospheric Habitability}

\TitleCitation{Multiverse Predictions for Habitability: Stellar and Atmospheric Habitability}

\Author{McCullen Sandora $^{1,}$*,
Vladimir Airapetian $^{2,3}$,
Luke Barnes $^{4}$ and
Geraint F. Lewis $^{5}$}

\AuthorNames{McCullen Sandora,
Vladimir Airapetian,
Luke Barnes and
Geraint F. Lewis}

\AuthorCitation{Sandora, M.; Airapetian, V.; Barnes, L.; Lewis, G.F.}

\address{
$^{1}$ \quad Blue Marble Space Institute of Science, Seattle, WA 98154, USA\\
$^{2}$ \quad Sellers Exoplanetary Environments Collaboration, NASA Goddard Space Flight Center, Greenbelt, MD, 20771, USA\\
$^{3}$ \quad Department of Physics, American University, Washington, DC, USA\\
$^{4}$ \quad School of Science, Western Sydney University, Locked Bag 1797, \mbox{Penrith South DC, NSW 2751, Australia}\\
$^{5}$ \quad Sydney Institute for Astronomy, School of Physics, A28, The University of Sydney, NSW 2006, Australia}

\corres{Correspondence: mccullen@bmsis.org}

\abstract{Stellar activity and planetary atmospheric properties have the potential to strongly influence habitability.  To date, neither have been adequately studied in the multiverse context, so there has been no assessment of how these effects impact the probabilities of observing our fundamental constants.  Here, we consider the effects of solar wind, mass loss, and extreme ultra-violet (XUV) flux on planetary atmospheres, how these effects scale with fundamental constants, and how this affects the likelihood of our observations.  We determine the minimum atmospheric mass that can withstand erosion, maintain liquid surface water, and buffer diurnal temperature changes.  We consider two plausible sources of Earth's atmosphere, as well as the notion that only initially slowly rotating stars are habitable, and find that all are equally compatible with the multiverse.  We consider whether planetary magnetic fields are necessary for habitability, and find five boundaries in parameter space where magnetic fields are precluded.  We find that if an Earth-like carbon-to-oxygen ratio is required for life, atmospheric effects do not have much of an impact on multiverse calculations.  If significantly different carbon-to-oxygen ratios are compatible with life, magnetic fields must not be essential for life, and planet atmosphere must not scale with stellar nitrogen abundance, or else the multiverse would be ruled out to a high degree of confidence.}


\begin{document}

\section{Introduction}

The multiverse hypothesis, which posits that other universes with different laws of physics exist, is an intriguing idea in theoretical cosmology that has so far proven challenging to test \citep{linde2017brief}.  This paper is part of a broader series aiming to rectify this, by generating a plethora of predictions within the multiverse framework regarding the nature of habitability \citep{mc1,mc2,mc3,mc4,mc5,mc6}. The core of this process is the requirement that the multiverse can only be a consistent theory of cosmology if it predicts that our presence in this particular universe is not too improbable; one way of falsifying the multiverse is to find that it predicts that the vast majority of complex (multicellular) life exists in universes with features different from our own.  Our contribution to this procedure lies in the recognition that the distribution of complex life, and so observers, throughout the multiverse, depends heavily on the assumptions we make about the nature of habitability.  Thus, certain habitability conditions, that are otherwise quite widely discussed, are incompatible with the multiverse.  If we ultimately find that the requirements for complex life are incompatible with the multiverse, we will be able to falsify the theory, to a calculable level of statistical significance.  Conversely, if we ultimately determine that all currently unknown habitability conditions turn out to be in line with multiverse expectations, we will accrue a long list of supporting evidence for the theory.

It remains to check the compatibility of each habitability condition with the multiverse framework by systematically incorporating them into our calculation of the distribution of observers throughout the multiverse, and the subsequent calculation of the probability of our observations.  To this end, we have organized this endeavor into several papers on the topic, each dealing with a loosely overarching theme.  The current paper explores several aspects relating to properties of planetary atmospheres, and stellar activity.  The two are tightly related, and considered by many to be essential for the maintenance of planetary habitability.

The compatibility of a habitability condition $\mathbb H$ with the multiverse is determined by the probability of observing our values of the fundamental constants.  This is communicated through the Bayes factor, which is defined relative to the baseline case where atmospheric effects are not important $\mathbb H_0$ by $\mathcal B(\mathbb H)= B(\mathbb H)/B(\mathbb H_0)$, where
\beq
B(\mathbb H)=\mathbb P(\alpha|\mathbb H)\,
\mathbb P(\beta|\mathbb H)\,
\mathbb P(\gamma|\mathbb H)\,
\mathbb P(\delta_u|\mathbb H)\,
\mathbb P(\delta_d|\mathbb H)
\eeq
and $\mathbb P(x|\mathbb H)=\min(P(x<x_\text{obs}|\mathbb H),P(x>x_\text{obs}|\mathbb H))$, for the fine structure constant $\alpha$, the electron to proton mass $m_e/m_p = \beta$, the proton to Planck mass $m_p/M_{pl} = \gamma$, the up quark to proton mass $m_u/m_p = \delta_u$, and the down quark to proton mass $m_d/m_p = \delta_d$.  The probability of observing particular values of the constants is defined through the probability density function $p(x|\mathbb H)\,\propto\, p_\text{prior}(x)\,\mathbb H(x)$, as described in more detail in \citep{mc1}.

For the baseline habitability condition $\mathbb H_0$, we take the most successful account of our observations we have considered, which is that complex life requires light from a relatively narrow spectral band for photosynthesis, and that the habitability of a planet is directly proportional to the amount of entropy it receives from incident starlight \citep{mc1,mc3}.

In Section \ref{Properties}, we discuss generalities of stellar properties, and how these vary with physical constants, deriving expressions that will be crucial for the rest of the paper.  In Section \ref{Atmosphere}, we discuss atmospheric loss processes, focusing in particular on extreme ultra-violet (XUV)-driven energy limited escape.  Determining the importance of this process as constants vary necessitates determination of a great many factors, including stellar spin-down history, initial atmospheric mass, and mass required for surface water retention, which we detail within.  Section \ref{B} is dedicated to stellar wind stripping present on planets without an intrinsic magnetic field, and the conditions for planetary magnetic fields to arise.

We find that the significance of atmospheric properties depends on which additional habitability assumptions are made.  If we take that an Earth-like carbon-to-oxygen ratio is required for life, as is commonly assumed, then the atmospheric conditions we consider do not strongly affect the probabilities we compute, and so they are neither favored nor disfavored by the multiverse.  However, if we adopt the stance that life does not depend on the carbon-to-oxygen ratio, several atmospheric conditions do strongly affect the multiverse probabilities.  Both the idea that atmospheric mass scales linearly with stellar nitrogen abundance and the idea that planetary magnetic fields are required for habitability cause the probability of our observations to significantly drop, and so both these conditions are incompatible with the multiverse hypothesis.  The strategy to test the multiverse is then to check whether this prediction is correct; if life indeed does not depend on planetary carbon-to-oxygen ratio, but either of these other two conditions is found true, the multiverse will be ruled out to high significance.

\section{How Do Stellar Properties Change in Other Universes?}\label{Properties}

Changes in stellar properties were among the first aspects to be investigated within a multiverse framework.  Refs. \citep{carter1974large,BT,carr1979anthropic} worked out how the properties of stars such as mass, lifetime, and luminosity change when constants vary.  Ref. \citep{schellekens} discuss photosynthetic potential of starlight.  Much work has been done on how different nuclear stability thresholds affect stellar fusion: Refs. \citep{dyson1971energy,bradford2009effect,barnes2015binding} investigated the effects of diproton stability.  \mbox{Refs. \citep{livio1989anthropic,oberhummer2000stellar,epelbaum2013dependence,huang2019sensitivity}} discuss the effects of alpha burning.  Refs. \citep{davies1983anthropic,BLdeuteron,adams2017habitability} investigate deuteron stability.  Ref. \citep{gould2012tritium} investigated the consequences of tritium stability.  Refs. \citep{adamsstars,howe2018nuclear} discuss non-nuclear energy production pathways.  Ref. \citep{BOreach} discuss the sizes of white dwarfs and neutron stars.  \mbox{Refs. \citep{adamsstarsandplanets,mc1}} discuss the entropy production as a key to habitability.

However, all of these previous studies have so far neglected some of the finer-grained stellar properties, which may nevertheless be just as important for determining the habitability of a planetary system.  Among these are properties of stellar coronae, magnetic fields, Sunspot fraction, stellar wind, rotation, and X-ray luminosity.  In part, this neglect may be due to prudence on the previous authors' parts, as many of these aspects remain imperfectly understood theoretically, making extrapolation of their behaviors to different universes fraught with potentially misplaced certainty.  However, much progress has been made in the understanding of many of these aspects in recent years, and we take advantage of these recent advances to establish a first attempt at determining how these properties may differ in other universes.

\subsection{Stellar Properties}
Expressions for stellar mass, radius, temperature, luminosity, and lifetime in terms of fundamental constants are all already well known (see, e.g., \citep{PL}), so we merely reproduce them here:
\bea
M_\star &=& 122.4\,\frac{\lambda\,M_{pl}^3}{m_p^2}\nonumber\\
R_\star &=& 108.6\,\frac{\lambda^{4/5}\,M_{pl}}{\alpha^2\,m_p^2}\nonumber\\
T_\star &=& 0.014\,\frac{\lambda^{19/40}\,\alpha^{1/2}\,m_e^{1/2}\,m_p^{3/4}}{M_{pl}^{1/4}}\nonumber\\
L_\star &=& 9.7\times10^{-4}\,\frac{\lambda^{7/2}\,m_e^2\,M_{pl}}{\alpha^2\,m_p}\nonumber\\
t_\star &=& 110.0\,\frac{\alpha^2\,M_{pl}^2}{\lambda^{5/2}\,m_e^2\,m_p}
\eea

The symbol $\lambda=M_\star/M_\text{Ch}$ is a dimensionless parameterization of stellar mass in terms of the Chandrasekhar mass $M_\text{Ch}=122.4M_{pl}^3/m_P^2=1.4M_\Sun$.  In these and all following expressions, the functional dependence on constants is derived using physical arguments, and the coefficients are set to accurately reproduce the correct values for our Sun, for the observed values of the physical constants. 

In addition, we will need the following expressions for the mass, density, orbital location, total incident power, and day length of an Earth-like planet, which is defined as both temperate (can maintain liquid surface water) and terrestrial (can retain heavy but not light atmospheric gases):

\bea
M_\text{terr} &=& 92\, \frac{\alpha^{3/2} \,m_e^{3/4}\,M_{pl}^3}{m_p^{11/4}}\nonumber\\
\rho_\text{rock}&=&0.13 \,\alpha^3 \,m_e^3 \,m_p\nonumber\\
a_\text{temp} &=& 7.6\, \frac{\lambda^{7/4}\,m_p^{1/2}\,M_{pl}^{1/2}}{\alpha^5 \,m_e^2}\nonumber\\
Q_\text{solar} &=& 5.3\times10^{-5}\,\frac{\alpha^7\,m_e^{9/2}\,M_{pl}^2}{m_p^{9/2}}\nonumber\\
t_\text{day} &=& 376\,\frac{M_{pl}}{\alpha^{3/2}\,m_e^{3/2}\,m_p^{1/2}}
\eea

Though there will be a certain tolerable range for each of these parameters, we specify to the Earth's values for our calculations.  Additionally, note that the temperate requirement has dictated that the incident stellar power is evaluated at $a_\text{temp}$ ($=$AU for our values), making this quantity independent of stellar mass.

\subsubsection{Speed of Stellar Wind}
The escape velocity of a star is 
\beq
v_\text{esc}=\sqrt{\frac{2\,G\,M_\star}{R_\star}}=0.30\,\lambda^{1/10}\,\alpha\label{vwind}
\eeq

For the Sun, this is 618 km/s.  The speed of solar wind is around 400--1000 km/s, roughly the same order of magnitude.  This results from the fact that the escaping wind is nonthermal, as particles that have enough energy to make it off the Sun usually have a surplus of the same order.  

This is larger than the thermal sound speed, which invariably depends on height.  For the photosphere, 
\beq
c_s\sim\sqrt{\frac{T_\star}{m_p}}=0.12\,\lambda^{19/80} \,\alpha^{1/4}\,\beta^{1/4}\,\gamma^{1/8}
\eeq

The sound speed of the corona is higher, as discussed below.

\subsubsection{Scale Height}
The scale height of a star is given by a competition between thermal and gravitational processes as 
\beq
H_\star\sim\frac{c_s^2}{g}=19.7\,\frac{\lambda^{43/40}m_e^{1/2}\,M_{pl}^{3/4}}{\alpha^{7/2}\,m_p^{9/4}}
\eeq

This is 100--1000 km for the Sun, and sets the scale for many processes, including the granule size and typical magnetic flux tube length.

\subsubsection{Stellar Magnetic Field}
The magnetic field at the stellar surface is created by a highly complex and incompletely understood dynamo mechanism \citep{brandenburg2005astrophysical,fan2009magnetic}.  However, the details of the precise mechanism are unimportant for determining the overall field strength, which is set by equipartition of energy as \citep{saar1986new} 
\beq
B_\text{surf}\sim\sqrt{4\pi\,P_\text{photosphere}}=3.1\times 10^{-5}\,\frac{\lambda^{19/20}\,\alpha\, m_e\,m_p^{3/2}}{M_{pl}^{1/2}}
\eeq

For the photosphere pressure, we use $P_\text{photosphere}\sim T_\star^4$, as appropriate for a $n=3$ polytrope, which describes stellar structure well \citep{kippenhahn1990stellar}.  The numerical value matches the observational quantity $B_\text{surf}\sim 2 G$.  This yields an estimate for the total field strength at the surface, which consists of both open field lines that contribute to the star's long range magnetic field, as well as highly complex field configurations that do not.  The long range field is related to the total strength by $B_\star= f_\text{open}B_\text{surf}$, where $f_\text{open}$ is the fraction of field lines which are open.  It is this factor that introduces rotational dependence to the stellar magnetic field.

\subsubsection{Fraction of Open Field lines}
The fraction of stellar magnetic field lines which are ``open'' (i.e., extend to infinity, rather than form a closed loop) depends both on stellar rotation and temperature.  This was postulated to depend on Rossby number in \citep{montesinos1993magnetic} as 
\beq
f_\text{open}=0.55\,\exp\left(-2.03\,Ro\right)
\eeq
\textls[-25]{where Rossby number is the ratio of rotation period to convective turnover time, \mbox{$Ro=P_\text{rot}/\tau_\text{conv}$}.
For the convective turnover time, we use the expression from \citep{gunn1998rotation}:}
\beq
\tau_\text{conv}=\tau_0\, \exp\left(-\frac{T}{T_\text{conv}}\right)\label{tauconv}
\eeq
where we have neglected terms that cause shutoff for large temperatures.  The turnover temperature is set by molecular absorption processes, $T_\text{conv}=0.27\alpha^2 m_e^{3/2}/m_p^{1/2}$.  
This is normalized to yield a Rossby number of 1.96 and an open field line fraction of 0.01 for the Sun.  The coefficient $\tau_0$ is set dimensionally to be $\tau_0\sim R_\star\sqrt{m_p/T_\star}=1.9\times10^5\lambda^{9/16}M_{pl}^{9/8}/(\alpha^{9/4}m_e^{1/4}m_p^{15/8})$, and is normalized to be 246.4 days for our Sun.  Expressing this in terms of fundamental parameters depends on the distribution of stellar rotation periods, which is discussed below.

\subsection{Corona}

The corona is the hotter, much less dense outer layer of a star.  Its properties are continuous with the star's extended stellar wind region of influence, and is the source region of most of the variable activity leading to space weather.

\subsubsection{Density of Corona}
In the formalism of \citep{cranmer2011testing}, the density of the corona (at the transition region) is determined by the equilibration of heating and cooling processes.  The heating rate is given by $Q_\text{heat}\sim \rho_\text{corona} v^3/\lambda_c$, where $\lambda_c$ is the granular scale, roughly set by the scale height $H=c_s^2/g$.  The cooling rate for bremsstrahlung is $Q_\text{cool}=n_e\,n_p\,\sigma_T\,v\epsilon$, where $\sigma_T=8\pi/3 \alpha^2/m_e^2$ is the Thomson cross section and $\epsilon\sim \alpha m_e$ is the typical energy \mbox{exchange \citep{carr1979anthropic}}.  These are equal when
\beq
\rho_\text{corona}\sim \frac{m_e\,m_p\,g}{\sigma_T\,\epsilon}=4.7\times10^{-7}\,\frac{\alpha\,m_e^2\,m_p^3}{\lambda^{3/5}\,M_{pl}}
\eeq

This is equal to $10^{-16}$ g/cm$^3$ for the Sun.

\subsubsection{Temperature of Corona}
The corona is about two orders of magnitude hotter than the photosphere, which has proven puzzling to explain theoretically for many years.  Consequently, various competing theories have been developed to explain the anomalously high temperature \citep{aschwanden2006physics}.  Perhaps the most popular account is that of Alfv\'en wave heating, which posits that energy is transferred to the corona from the stellar interior by turbulent plasma oscillations.  In the following, we only consider this theory, which gives the heat flux as \citep{malara2001observations}:
\beq
S=\frac12\,\rho_\text{corona}\, \delta v^2\, v_A
\eeq

Here $\delta v^2\sim T/m_p$ and $v_A=B/\sqrt{\rho_\text{corona}}$.  This determines temperature through the diffusion equation $S=-\kappa_\text{th}\nabla T\sim \kappa_\text{th} T/H_\star$.  From \citep{brandenburg2005astrophysical}, the thermal conductivity of a stellar plasma is
\beq
\kappa_\text{th}\sim \frac{1.31}{\pi\log\Lambda}\frac{T^{5/2}}{e^4\,m_e^{1/2}}
\eeq
where $\log\Lambda\sim$5--20 is the Coulomb logarithm, which has mild parameter dependence, but can be ignored.  This can be solved for $T$ to yield
\beq
T_\text{corona}\sim \left(\frac{e^4\,m_e^{1/2}}{m_p^2}\,\frac{\rho_\text{corona}^{1/2}\,B_\text{surf}}{g}\right)^{2/3}=4.6\times10^{-3}\,\frac{\lambda^{5/6}\,m_e^{5/3}}{\alpha^{1/3}\,m_p^{2/3}}
\eeq

\subsection{Stellar Wind}

\subsubsection{Mass Loss Rate}
According to \citep{kippenhahn1990stellar}, many analytic mass loss formulas have no strong theoretical justification.  Whatever the underlying mechanism for solar wind, it is constrained by the continuity equation to obey
\beq
\dot M \sim \rho_\text{corona}\, v\, 4\pi\, R_\star^2 = 7.0\times10^{-5}\,\frac{\lambda^{11/10}\,m_e^2\,M_{pl}}{\alpha^2\,m_p}
\eeq

This is normalized to yield $2\times10^{-14}M_\Sun$/yr for the Sun.  With this, we may ponder whether in some universes the stellar wind is strong enough to deplete stellar material before the available nuclear energy is exhausted; in such universes, type II supernovae would not occur, with stars instead ending their lives having blown off material to the point where fusion ceases.  We find $t_\star \dot M/M_\star=.16\alpha^5\beta^{1/4}/(\lambda^{3/2}\gamma)=3\times10^6$, so that if $\alpha$ were a factor of 20 lower, this would indeed be the case.  However, this may not preclude the distribution of heavy elements into the interstellar medium, if enough reach the wind-launch site.  More work is needed to determine whether this mechanism can be at play, whether heavy elements collect in the stellar core, or whether the strong wind effectively extinguishes the star before any heavy elements are created. In any case, including this boundary in parameter space does not appreciably affect the probabilities we compute.

\subsubsection{Alfv\'en Radius}
The Alfv\'en radius is the point at which an appreciable azimuthal velocity component develops.  This is set by 
\beq
\frac{B_\star^2}{4\pi}\sim \rho(r) v_r^2
\eeq

Throughout we take the Sun's Alfv\'en radius to be $R_A\sim 24 R_\Sun$, though it can vary by a factor of 2 throughout the solar cycle \citep{goelzer2014analysis}.
By the continuity equation, the quantity $\rho\, v_r\propto1/r^2$.  The radial dependence of $v_r$ can be found using Parker's model of solar wind, which gives
\beq
\frac{1}{v_r}\left(v_r^2-c_s^2\right)\frac{dv_r}{dr} = \frac{2c_s^2}{r}-\frac{G\,M_\star}{r^2}
\eeq

If we define the sonic radius $R_s=GM_\star/(2c_s^2)$, then for $r\gg R_s$, this gives $v_r\rightarrow 2c_s\log(r/R_s)^{1/2}$ \citep{ryden2011radiative}, though to first approximation the logarithmic dependence can be neglected.

If $B$ is primarily dipolar, $B(r)=B_\star(R_\star/r)^3$, and we find
\beq
R_A\sim\left(\frac{f_\text{open}^2\,T_\star^4}{\rho_\text{corona}\, c_s^{2}}\right)^{1/4}\,R_\star=1.1\times10^4\,\frac{\lambda^{11/8}\,f_\text{open}^{1/2}\,M_{pl}}{\alpha^{9/4}\,m_p^2}
\eeq

For more generic magnetic field profiles $B\sim r^{-q}$, the fourth root is replaced by $1/(2q-2)$.

\subsubsection{X ray Luminosity}
A star's X-ray luminosity, which is an important driver of planetary atmospheric loss, is greatly enhanced with respect to the thermal contribution by dynamo processes.  As such, X-ray luminosity is found to correlate well with both magnetic activity and rotation speed for slowly rotating stars \citep{wright2011stellar}.  For stars with rotation periods less than a few days, however, the X-ray luminosity is found to saturate to about $10^{-3}$ of the bolometric luminosity.  The origin of this is not well understood, but could be due either to the saturation of surface magnetic flux, or internal dynamo \citep{pizzolato2003stellar}, representing a qualitatively different regime of energy transport.  These two regimes can be encapsulated with the following expression

\beq
L_X=\frac18\,B_\star^2\,R_\star^2\,\min(v_\text{conv},v_\text{rot})\label{Lx}
\eeq
which reproduces the linear rotation-activity relation between X-ray luminosity and magnetic flux found in \citep{pevtsov2003relationship}.  Here, we have defined a convective speed in terms of the convective turnover time in Equation (\ref{tauconv}) as $v_\text{conv}=R_\star/\tau_\text{conv}$.

\subsection{Rotation}

Since stellar activity depends on rotation rate, and stellar rotation decreases over time, the majority of a planet's atmospheric loss may occur during the initial phase of stellar evolution.  Here, we derive expressions for initial stellar rotation as well as spindown rate.

\subsubsection{Initial Stellar Rotation}
Stars are observed to have a spread of rotation periods within the span of several days, with periods that increase as they age \citep{gallet2013improved}.  At formation time, one may expect that stars inherit their rotation from their collapsed dust cloud, but an order of magnitude estimate reveals that the angular momentum of the dust cloud vastly exceeds stellar angular momentum \citep{bodenheimer1995angular}.  Indeed, a star possessing that much angular momentum would exceed the critical breakup velocity, and would quickly jettison its material.  Instead, the star radiates angular momentum through its surrounding disk until it drops below the breakup speed, and can coalesce \citep{gallet2013improved}.  This process results in initial stellar rotation frequencies being close to their breakup velocity, as observed in \citep{kawaler1987angular}:
\beq
\Omega_0= \sqrt{\frac23\frac{G\, M_\star}{R_\star^3}}=1.6\times10^{-3}\,\frac{\alpha^3\,m_p^2}{\lambda^{7/10}\,M_{pl}}
\eeq

\subsubsection{Stellar Spindown Time}
Stars lose angular momentum throughout their evolution via stellar wind.  While a star's angular momentum is given by  $J\sim M R_\star^2\Omega$, to estimate angular momentum loss we must keep in mind that the stellar wind travels radially outward until the Alfv\'en radius, and so angular momentum loss is given by $\dot{J}\sim \dot{M} R_A^2 \Omega$ \citep{weber1967angular}.  This increased lever arm greatly enhances spindown, and also introduces extra rotation dependence, as the Alfv\'en radius depends on spin.  A linear dependence $R_A\propto \Omega$ leads to a cubic evolution equation for $\Omega$, as first discussed in \citep{durney1977angular}.  Additionally, a qualitative shift in spindown behavior empirically occurs when the rotation frequency exceeds a critical value, akin to the convective turnover time given in Equation (\ref{Lx}).  This leads to the following equation governing the evolution of rotation \citep{gallet2013improved}:
\beq
\dot \Omega
\sim-\frac{B_\star^2\,R_\star^2}{M_\star\,v}\,\Omega\,\min(\Omega^2\,\tau_\text{conv}^2,1)\label{Omegadot}
\eeq

This also sets the spindown time as 
\beq
t_\text{brake}\sim\frac{M_\star}{\dot M}\frac{R_\star^2}{R_A^2}=.21\,\frac{\alpha^{5/2}\,M_{pl}^2}{\lambda^{5/4}\,f_\text{open}\,m_e^2\,m_p}
\eeq

For stars rotating more rapidly than the convective turnover time, spindown is set by the star's convective churn, rather than rotation.  Below this, the evolution $\dot\Omega\sim\Omega^3$ leads to the well established Skumanich law, $P_\text{rot}\sim\sqrt{t}$ \citep{skumanich1972time}.  For fast rotators, the decay is instead exponential.

These are all the properties of stars we will need to model our habitability effects in the sections below.

\section{How Do Atmospheric Properties Differ in Other Universes?}\label{Atmosphere}
We now turn our attention to planetary atmospheres, and whether their character is substantially different in other universes.  In particular, we ask what physics determines that the Earth's atmospheric mass is six orders of magnitude less than the planet's mass, how this compares to the minimum needed for several habitability considerations, and whether the expected atmospheric mass is lower than these thresholds for different values of the fundamental constants.

At first glance, it may seem strange to attempt to explain atmospheric mass fraction in terms of fundamental constants.  After all, the solar system alone exhibits an enormous diversity of atmospheric mass fractions amongst its planets, from almost zero around small inner rocky bodies to nearly unity for the gas giants.  Indeed, atmospheric mass seems to depend on a great number of variables: planetary mass, interior and surface chemistry, orbit, evolution, flux, and the presence or absence of life \citep{stueken2016modeling}.  Even Venus, though remarkably similar to Earth in orbit and mass, has an atmosphere 90 times Earth's.  However, closer inspection reveals hidden regularity; Venus's atmospheric nitrogen content is only \mbox{3--4 times} that of Earth's, placing it at the same order of magnitude \citep{wordsworth2016atmospheric}.  Its carbon dioxide content, which comprises the bulk of the atmosphere, is the same order of magnitude as that found dissolved in Earth's oceans and compressed into sedimentary rock \citep{ingersoll2013planetary}.  Even Venus's initial water content is estimated to have been similar to Earth's \citep{ingersoll2013planetary} {(though recent work indicates that even if its initial water content were similar, it may not have ever been able to condense from a steam atmosphere to form an ocean} \citep{turbet2021day}).  Evidently, this diversity stems from the different phases each species can undergo, rather than the primordial abundance of each element, giving hope that the overall mass fraction may be understood by processes operating in the early solar system, as well as galactic element abundances.  Furthermore, if this is the case, we have hope of extrapolating these values to other universes.

In the following, we focus on nitrogen, as the only gas which is noncondensible under temperate conditions, and present in appreciable quantities.  Its presence is essential for stability of liquid water on Earth's surface \citep{wordsworth2013water}.  It was estimated in \citep{johnson2015nitrogen} that the nitrogen contained in the Earth's mantle is between 3--10 times that of Earth's atmospheric nitrogen, a ratio that is certainly affected by the presence of other species, but is likely to hold as a rough order of magnitude estimate under a range of conditions \citep{wordsworth2016atmospheric}.  Earth's atmospheric nitrogen has remained constant to within a factor of two over the past 3 Gyr, as evidenced by analyzing raindrop imprint size \citep{som2012air} and the isotopic composition of quartz \citep{marty2013nitrogen}.

In the following, we consider two explanations for the magnitude of Earth's nitrogen abundance, corresponding to two different plausible sources: late accretion by chondrites, and initially, as dissolved material in Earth's original building blocks.  Each of these hypotheses has different implications for the amounts of nitrogen on planets elsewhere in our universe, as well as throughout the multiverse.  Additionally, it is an open question how planetary nitrogen abundance scales with initial stellar system nitrogen abundance, which has important implications for the multiverse, as we happen to be very close to a boundary beyond which nitrogen abundance is reduced by a factor of 270.  At the two extremes, the dependence may be linear, if the nitrogen content of solar system bodies was not close to their carrying capacity, or independent, if the bodies were saturated.  The dependence probably lies somewhere between these two extremes, but we report how adopting each assumption alters our multiverse probabilities, which serves to bracket the upper and lower limits for our calculations.

We then consider three atmospheric mass thresholds that are plausibly related to habitability.  The first is the amount of atmosphere that can be stripped away by stellar flux.  The second is related to the pressure necessary to maintain liquid surface water.  Third, the mass needed to buffer diurnal temperature changes.  Finally, we consider the possibility that only initially slowly rotating stars in our universe are capable of retaining their atmospheres, and assess the compatibility of this hypothesis with the multiverse.

\subsection{Possible Sources of Atmosphere}

The fact that Earth possesses an atmosphere containing volatile constituents is somewhat of a mystery, given that the conditions during Earth's formation were much hotter than their condensation temperatures.  Naively, this would result in inner planets that are almost completely comprised of refractory elements, which is manifestly not the case.  In the following, we consider two leading theories for the origin of Earth's nitrogen atmosphere: delivery during late accretion from outer system bodies, and as a result of initial accretion from nitrogen dissolved in Earth's original building blocks.

\subsubsection{Initial Atmosphere Delivered during Accretion}
The classic account for Earth's volatile budget is from planetesimals initially situated outside the solar system's ice line, where temperatures were below the condensation point of volatile species.  It has been estimated that up to 7.5 atmospheric masses could have been delivered by carbonaceous chondrites after the main phase of planet formation was completed \citep{javoy1997major}.  This account has the simplicity of explaining the origin of Earth's atmosphere and ocean by a single common source.  Additionally, it can readily explain the hierarchy of why Earth's ocean is $\sim$100 times more massive than the atmosphere, \mbox{as \citep{schaefer2010chemistry}} demonstrated that the H$_2$O/N$_2$ impact degassing ratio is $\sim$100 for a range of different chondrites.  Finally, we would like to stress that in this scenario, final atmospheric mass will be highly stochastic, as the material delivered through late accretion is dominated by few large bodies \citep{sinclair2020evolution}.  Thus, while we compute the expected value, it should be kept in mind that this scenario yields a distribution of atmospheric mass ratios.

In \citep{mc6}, we derive the planetary ocean mass fraction delivered via planetesimal accretion during planet formation in terms of the amount of material delivered during late accretion.  In this scenario, the atmospheric volatiles are delivered in the same manner.  Therefore, we may posit the atmospheric mass fraction to simply be
\beq
f_N=0.011\,\frac{\kappa\,\lambda^{21/10}\,\gamma^{1/3}}{\alpha^{11/2}\,\beta^{25/12}}\label{fN1}
\eeq

For details on how this expression was obtained, we refer the reader to \citep{mc6}.

\subsubsection{Atmospheric Mass as a Result of Accretion by N-Rich Bodies}
Here, we follow \citep{grewal2021rates} by considering that Earth's nitrogen was delivered during accretion in the form of dissolved N inside rock and metal.  We may then derive the total amount of resulting nitrogen as a function of body mass, with the presumption that only nitrogen in the interior of these planetesimals will be incorporated into the planet's final budget.  

The initial nitrogen fraction of a planetesimal is $f_N=m_N/m_\text{pp}$, where $m_{pp}$ is the mass of the planetesimal.  The resultant nitrogen budget is obtained through the magma ocean and core as
\beq
\hat f_N=\frac{m_N^{MO}+M_N^\text{core}}{m^{MO}+M^\text{core}}=\frac{1+Z\, D_N}{1+Z}C_N^{MO}
\eeq
where $Z=M^\text{core}/M^\text{mo}$, $C_N^\text{MO}=M_N^\text{MO}/M^\text{MO}$, and $D_N=C_N^\text{core}/C_N^\text{MO}$.

For the fraction of nitrogen dissolved in the magma ocean, we use \citep{libourel2003nitrogen}:
\beq
C_N^\text{MO}=\frac{p_N}{p_1}+fO_2^{-3/4}\,\left(\frac{p_N}{p_2}\right)^{1/2}
\eeq
where $p_1$ and $p_2$ are coefficients, taken here to scale as $p_i\propto Ry^4$, with $Ry$ the Rydberg constant that dictates the electronic energy scale.  The quantity $fO_2$ is oxygen fugacity, and will depend of the primordial abundances of the two elements.

The partial pressure can be rewritten in terms of atmospheric nitrogen mass as 
\beq
p_N=M_N^\text{atm}\,\frac{g}{A}=(M_N^\text{tot}-M_N^\text{MO}-M_N^\text{core})\frac{g}{A}=\frac{M_{pp}\,g}{A}\left(f_N-(1-f_N)\hat f_N\right)
\eeq

This can then be used to find an equation determining $\hat f_N$:
\beq
\hat f_N=k_1(f_N-\hat f_N)+\sqrt{k_2(f_N-\hat f_N)}
\eeq
where for cleanliness we have defined $k_1=\zeta\tau/p_1$, $k_2=\zeta^2 fO_2^{-3/2}\tau/p_2$, $\zeta=(1+Z D_N)/(1+Z)$, and $\tau=gM_{pp}(1-f_N)/A$.  This can be solved for $\hat f_N$ to find  
\beq
\hat f_N = \frac{2f_Nk_1(1+k_1)-k_2+\sqrt{4f_N(1+k_1)k_2+k_2^2}}{2(1+k_1)^2}
\eeq

We find that for large mass bodies, $k_1$,$k_2\rightarrow\infty$, $\hat f_N\rightarrow f_N$, so that planetary nitrogen abundance matches the primordial value.  In the limit $k_2\rightarrow0$, this expression simplifies significantly to $\hat f_N\rightarrow f_N k_1/(1+k_1)$.  This expression allows us to derive the final nitrogen abundance as a function of planetesimal mass, by noting that $g/A\sim G \rho_\text{rock}^{4/3}/M^{1/3}$, $D_N\sim \text{exp}((b+cP)/T)$, $T\sim GM/R\sim GM^{2/3}\rho_\text{rock}^{1/3}$.

To determine the planetary nitrogen abundance fraction that results from original accretion, we need the typical planetesimal size.  For this, we use the isolation mass $M_\text{iso}=1.3\times 10^8\,\kappa^{3/2}\,\lambda^{25/8}\,m_p^{7/4}\,M_{pl}^{9/4}/(\alpha^{15/2}\,m_e^3)$ \citep{mc2}.  In the limit that $k_1\ll1$ and neglecting the dependence on planetary mass of $D_N$, this gives
\beq
\hat f_N=7.6\times 10^{-7}\,\frac{\kappa\,\lambda^{25/12}\gamma^{1/2}}{\alpha^9\,\beta^2}
\eeq

Interestingly, the dependence on stellar mass of this quantity is practically indistinguishable from that of the alternate nitrogen source, Eqn. (\ref{fN1}).

\subsection{Which Atmospheric Thresholds Are Important for Habitability?}

Earth's atmosphere is quite comfortably above any catastrophic thresholds, being about two orders of magnitude larger than needed to prevent total atmospheric escape, maintain liquid surface water, and buffer diurnal temperature changes.  However, given the exponential dependence on constants of some of these conditions, we investigate the influence each exerts on our multiverse calculations.

\subsubsection{How Much Atmospheric Loss Occurs in Other Universes?}
In this paper, we restrict our attention to terrestrial planets, which are defined such that light gases such as hydrogen and helium, but not heavy gases such as water, oxygen and nitrogen, undergo Jeans escape.  For these planets, the dominant form of atmospheric escape is driven by stellar XUV light, and is in the energy limited regime (for recent reviews, see \citep{catling2017escape,gronoff2020atmospheric}).  The mass loss rate for this type of escape is given by equating the energy of UV light absorbed by the atmosphere with the energy of atmospheric particles ejected at the escape speed \citep{zahnle1990mass},
\beq
\dot M_\text{XUV}=\epsilon\,\frac{ R_\Earth^3\,L_\text{X}}{\,a_\text{temp}^2\,G\,M_\Earth}
\eeq

Here, $\epsilon$ is an unimportant efficiency factor.  This is independent of atmospheric mass, being limited by the amount of energy imparted in the upper atmosphere rather than the amount of material present.  In \citep{johnstone2019extreme} it was estimated that an XUV flux greater than $60$ times Earth's value would be needed to induce a catastrophic mass loss rate of $1.8\times10^9$ g/s, capable of eroding the entire atmosphere.  For reference, M dwarfs and young K dwarfs are subjected to 100--400 times Earth's XUV flux \citep{airapetian2020impact}.

The total atmospheric mass loss through X-ray flux may be found through \mbox{Equation (\ref{Lx})}, taking rotation evolution into account:
\beq
\Delta M_\text{XUV}\sim\frac{R_\Earth^3}{a_\text{temp}^2\,G\,M_\Earth}\frac{B_\star^2\,R_\star^3}{8}\int_0^{t_\star}dt\,\min(\Omega_\text{conv},\Omega(t))
\eeq

Using the evolution dictated by Equation (\ref{Omegadot}) and in the limit $t_\star\gg t_\text{brake}$, this integral can be performed to find
\beq
\Delta M_\text{XUV}\sim\frac{B_\star^2\,R_\star^3}{a_\text{temp}^2\,G\,\rho_\text{rock}}\,\min(\Omega_\text{conv},\Omega_0)\,\sqrt{t_\star\,t_\text{brake}}
\eeq

\begin{figure}[H]
		\includegraphics[width=.7\textwidth]{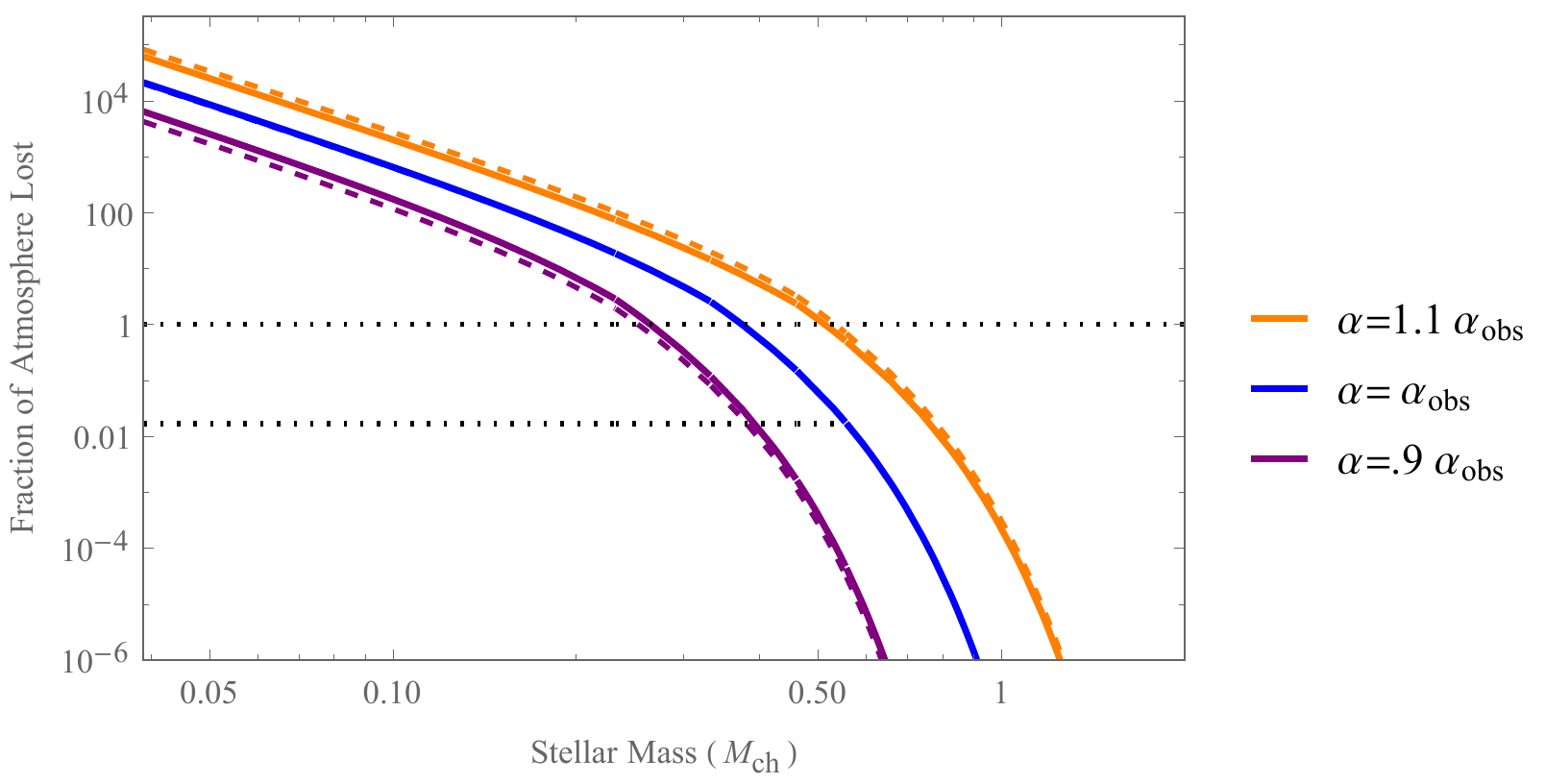}
		\caption{Fraction of atmosphere lost as a function of stellar mass.  The dependence of this quantity on the fine structure constant $\alpha$ can be observed.  The solid lines assume a late atmospheric delivery scenario, and the dashed lines assume atmosphere originates in original accretion.  The dotted lines correspond to the catastrophic value 1, and the Sun's value.}
		\label{fatmloss}
	\end{figure}

The condition $t_\star\gg t_\text{brake}$, which holds by three orders of magnitude in our universe, is not necessarily generic; we compute $t_\text{brake}/t_\star=0.0019\lambda^{5/4}\alpha^{1/2}/f_\text{open}$, 
which can be much larger than 1 if no stellar magnetic field lines are open for certain parameters.  However, including a more complete expression does not affect the calculated probabilities appreciably, while considerably complicating the formulae.  In Figure \ref{fatmloss}, we display the atmospheric mass loss for temperate, terrestrial planets as a function of stellar mass, for three different values of the fine structure constant.  The difference resulting from adopting the two alternate origin scenarios is also displayed, but is seen to be minimal.  This defines some stellar mass below which more than the initial atmosphere is lost through XUV irradiation, which depends on fundamental constants, and can be larger than the solar mass ($\lambda=1/1.8$) in some regions of parameter space.

\subsubsection{How Much Atmosphere Is Needed to Maintain Liquid Surface Water?}
Liquid surface water can exist only when atmospheric pressure exceeds that at the triple point, where the three low energy phases of water coexist in equilibrium.  The location of the triple point can be determined by noting that the solid-liquid transition is almost independent of pressure, and occurs at temperature set by the vibrational molecular energy $T_\text{freeze}\sim \alpha^{1/2}/(m^{1/2}r_{H_2O}^{3/2})$.  The liquid-gas transition is given by the Clausius--Clapeyron equation as $P(T)=P_0e^{-L/T}$, where the latent heat of evaporation is $L\sim \alpha r_{H_2O}$.  The coefficient $P_0$ can be found by enforcing that the phase curve terminates at the observed critical point of water of 647 K and 22.1 MPa.  Though an imperfect description, the van der Waals equation of state may be used to provide a theoretical expectation for the location of the critical point in terms of the molecular radius and energy $\epsilon$, yielding $T_\text{crit}=8/27\, \epsilon$ and $P_\text{crit}=\epsilon/(18\pi r^3)$ \citep{hill1986introduction}.  Normalizing to fit our observed values, this yields the pressure at the triple point to be
\beq
P_\text{triple}=\frac{\epsilon}{18\pi\,r^3}\, e^{\frac{27 L}{8\,\epsilon}}=1.6\times10^{-3}\,\alpha^5\,m_e^4\,e^{-.424/\sqrt{\beta}}
\eeq

This can then be related to minimal atmospheric mass capable of supporting liquid water through $M_\text{min}=4\pi R_\Earth^2 P_\text{triple}/g$, giving
\beq
M_\text{min}=0.87\,\frac{\alpha^{3/2}\,m_e^{1/4}\,M_{pl}^3}{m_p^{9/4}}\,e^{-.424/\sqrt{\beta}}
\eeq

This is about $0.006M_\text{atm}$ for our values.

\subsubsection{How Much Atmosphere Is Needed to Buffer Diurnal Temperature Changes?}
Earth's atmosphere retains substantial heat, which buffets the day--night temperature difference from the otherwise extreme variations that would occur, such as the day--night temperature differences on the Moon and Mars which can reach hundreds of degrees Kelvin.  This occurs because the relaxation time of Earth's atmosphere, estimated as the ratio of thermal energy over the power supplied, $t_\text{relax}\sim E_\text{therm}/Q_\text{solar}$, is about 100 days.  This gives
\beq
t_\text{relax}\sim \frac{T\,M_\text{atm}\,a_\text{temp}^2}{m_p\,L_\star\,R_\Earth^2}
\eeq

For small enough atmospheric mass, this is less than half a day, and the atmosphere does not play a significant role in averaging out daily variations of stellar flux.  This occurs at the threshold
\beq
M_\text{min}=1.9\,\frac{\alpha^{7/2}\,m_e^{3/2}\,M_{pl}^3}{m_p^{7/2}}
\eeq 

The exact mass depends strongly on water content, as evidenced by the extreme temperature differences present in Earth deserts, but we do not consider this here.

\subsubsection{Are Only Slowly Rotating Stars Habitable?}
There is evidence from noble gas isotopes \citep{odert2018escape}, the Moon \citep{saxena2019Sun}, and Venus \citep{lammer2021formation} that the Sun began as an anomalously slow rotator.  However, it is not currently possible to determine precisely how slow, and many studies only differentiate between stars in the lower 25 percentile.  If true, this suggests a selection effect: ordinarily rotating stars may be incapable of hosting life, presumably due to high early atmospheric loss.

To determine the compatibility of this habitability hypothesis with the multiverse, we follow the fraction of slowly rotating stars
\beq
f_\text{slow} = \min\left(\frac{M_\text{atm}}{\Delta M_\text{XUV}},1\right)
\eeq

This treats the initial rotation distribution as uniform up to the natural value $\Omega_0$, which is loosely consistent with observations of stellar populations \citep{wolff1982origin}.  To account for the observation that the Sun appears to be in the lower 25 percentile, we rescale the fraction of slow rotators (of Sun-like stars) in our universe to be 1/4.

\subsection{Is Atmospheric Stability a Factor Determining Our Presence in This Universe?}

We can now test the various atmospheric habitability thresholds, on the basis of their compatibility with our observations within the multiverse.  To this end, we test the following four thresholds: loss due to XUV radiation, the minimal mass for stable liquid surface water, the minimal mass to buffer diurnal temperature changes, and the notion that only stars which are slowly rotating are habitable.  In addition, we check both the early and late origin scenarios for our atmosphere, both an independent and linearly dependent abundance as a function of stellar nitrogen content, and either restricting to a narrow range of Earth-like carbon-to-oxygen values, or not.  In Table \ref{airtable}, we display the various Bayes factors for each of these combinations.

We find that when restricting consideration to a narrow range of carbon-to-oxygen ratios, the Bayes factors for the various habitability criteria do not vary significantly.  When considering the carbon-to-oxygen ratio to not play a factor in habitability, however, several of the habitability criteria are severely disfavored in the multiverse context.  The disfavored criteria all have to do with the assumption that planetary nitrogen content scales linearly with stellar system nitrogen abundance, and does not depend on the atmospheric source or threshold mass.  This is a consequence of our universe being situated very close to a precipitous threshold where nitrogen-14 is unstable \citep{mc5}, which affects the probabilities if the carbon-to-oxygen ratio is unimportant but does not if restricted to the subspace where the carbon-to-oxygen ratio is close to our observed value.  We note that in \citep{mc5} we found additional reasons to favor a restricted range of carbon-to-oxygen ratio based on the observed Hoyle energy value and organic to rock ratio in our universe.  Apart from this insight, no strong preference can be given to the different atmospheric origin scenarios, threshold masses, or expectation on whether only slow rotators are habitable.  Our conclusion is that it is certainly consistent that atmospheric mass may play a large role in the habitability of our universe, but it does not appear to be a driving factor in determining our particular observations.

\begin{table}[H]
	\caption{Bayes factors for various atmospheric habitability criteria relative to the baseline case where atmosphere mass is unimportant for habitability. Small values indicate that a set of assumptions is disfavored to a corresponding degree in the multiverse framework. The cases considered are that atmosphere must be large enough to withstand XUV loss, be above the triple point of water, can buffer diurnal temperature changes, and that only slowly rotating stars are habitable.  The late delivery vs. initial columns consider both potential origins of the atmosphere, and the N dep columns consider that planetary nitrogen abundance scales linearly with stellar system abundance.  The top rows restrict to Earth-like values of C/O ratio, and the bottom do not.}
	\label{airtable}
	\setlength{\cellWidtha}{\textwidth/5-2\tabcolsep+0.0in}
\setlength{\cellWidthb}{\textwidth/5-2\tabcolsep+0.0in}
\setlength{\cellWidthc}{\textwidth/5-2\tabcolsep+0.0in}
\setlength{\cellWidthd}{\textwidth/5-2\tabcolsep+0.0in}
\setlength{\cellWidthe}{\textwidth/5-2\tabcolsep+0.0in}
\scalebox{1}[1]{\begin{tabularx}{\textwidth}{>{\PreserveBackslash\centering}m{\cellWidtha}>{\PreserveBackslash\centering}m{\cellWidthb}>{\PreserveBackslash\centering}m{\cellWidthc}>{\PreserveBackslash\centering}m{\cellWidthd}>{\PreserveBackslash\centering}m{\cellWidthe}}

			\toprule
			\pmb{$\mathbb H$} & \textbf{Late Delivery} & \textbf{Late (N Dep)} & \textbf{Initial} & \textbf{Initial (N Dep)} \\
			\midrule
			\multicolumn{5}{l}{Earth-like C/O}\\
			\midrule
			$M_\text{atm}>\Delta M_\text{XUV}$ & 0.87 & 0.66 & 0.94 & 0.71\\
			$M_\text{atm}>M_\text{triple}$ & 0.82 & 0.63 & 0.89 &0.67\\
			$M_\text{atm}>M_\text{diurnal}$ & 1.0 & 0.75 & 1.0 & 0.75\\
			slow rotator & 0.32 & 0.25 & 0.40 & 0.32\\
			\midrule
			\multicolumn{5}{l}{Unrestricted C/O}\\
			\midrule
			$M_\text{atm}>\Delta M_\text{XUV}$ & 2.05 & 0.0041 & 2.34 & 0.0038\\
			$M_\text{atm}>M_\text{triple}$ & 1.09 & 0.0057 & 1.84 & 0.0032\\
			$M_\text{atm}>M_\text{diurnal}$ & 1.98 & 0.0074 & 3.08 & 0.068\\
			slow rotator & 1.14 & 0.00067 & 2.62 & 0.00082\\
			\bottomrule
		\end{tabularx}}

\end{table}

\section{Are Planetary Magnetic Fields Generic?}\label{B}
A planet's magnetic field is purported to be essential for habitability, as it shields against charged particles, preventing stellar wind stripping (see, e.g., \cite{lammer2007coronal}).  However, it must be pointed out that magnetic fields also provide several avenues for ion escape \citep{gunell2018intrinsic}, which may well represent the dominant form of atmospheric loss on Earth today \citep{ramstad2021intrinsic}.  Indeed, Venus has managed to retain its atmosphere without an intrinsic (as opposed to induced by the Sun's) magnetic field, despite being closer to the Sun than Earth.

Given the uncertain importance of planetary magnetic fields for habitability, we ask whether their properties change significantly in other universes, and thus whether demanding their presence influences the probabilities of any of our observables.  We focus on five relevant aspects required for a magnetic field to be both present and protective:
(i) The core's magnetic Reynolds number is large enough to support a dynamo.  (ii) The magnetosphere must extend beyond the atmosphere, as otherwise it will have little effect on loss properties.  (iii) The star's temperate zone must be outside its Alfv\'en zone, as otherwise the planetary and stellar magnetic field lines connect, forming a direct line of transport which dumps stellar wind onto the planet's poles, rather than act as a shield.  (iv) The development of a magnetic field requires a metal core, placing limits on the oxygen content of the planet.  (v) The magnetic field is generated through a dynamo, and so requires the core to remain at least partly liquid for an appreciable duration.  If planetary magnetic fields are essential for habitability, all of these conditions must be met.

\subsection{When Is the Magnetic Reynolds Number Large Enough to Induce a Dynamo?}
Both theory and simulations of the Earth's core indicate that a dynamo will only exist when advection of the magnetic field dominates over diffusion \citep{roberts2000geodynamo}.  This can be summarized as a condition on the magnetic Reynolds number $R_a=v_\text{core}\,L/\eta>$\,10--100 (Earth's magnetic Reynolds number is about $10^3$) \citep{davies2015constraints}.  This condition can be used to place constraints on the fundamental constants, using the length scale $L\sim R_\text{core}$, and magnetic diffusivity $\eta=1/(4\pi\sigma_\text{electric})$ with $\sigma_\text{electric}$ the electrical conductivity, which is related to thermal conductivity through the Wiedemann--Franz law  \citep{poirier2000introduction}:
\beq
\frac{\hat\kappa_\text{heat}}{\sigma_\text{electric}}=\frac{\pi^2}{3}\frac{T}{e^2}
\eeq

In \citep{mc3}, we found an expression for the thermal diffusivity in terms of fundamental constants as $\kappa_\text{heat}=2/(m_e^{1/4}m_p^{3/4})$, which is related to thermal conductivity through $\kappa_\text{heat}=\hat\kappa_\text{heat}/(c_p\rho_\text{rock})$.  The core convective speed can be obtained from mixing length theory, $v_\text{core}\sim (L q/(\rho_\text{rock} H_T))^{1/3}$ \citep{stevenson2003planetary}.  Using our expression for heat flux $q=0.58 \alpha^{11/2}m_e^5/M_{pl}$ \mbox{from \citep{mc3}} and the generic expression for scale height $H_T\sim c_s^2/g$, we find
\beq
R_a = 0.33\,\frac{\alpha^{7/3}\,\beta^{7/6}}{\gamma^{2/3}}
\eeq

These scalings are not significantly altered if we instead use the magnetostrophic estimate for the core convection velocity, also from \citep{stevenson2003planetary}.

\subsection{Is the Magnetosphere Always Larger Than the Atmosphere?}
In order for a planetary magnetic field to be an effective shield against stellar wind, it must extend beyond the atmosphere.  The size of the magnetosphere can be estimated as the point at which the magnetic pressure is equal to the stellar wind pressure, yielding for a dipole field \citep{kivelson2014planetary} the standoff distance:
\beq
r_\text{magnetosphere} = \left(\frac{2\,B_0^2}{\rho_\text{sw}\,v_\text{sw}^2}\right)^{1/6}
\eeq

To evaluate this, we use the expressions for density and speed of solar wind from Section \ref{Properties}.  It remains to estimate $B_0$, the strength of the magnetic field at the planet's surface.

There are an inordinate number of proposals for how planetary magnetic field strength depends on planetary characteristics, as reviewed in \citep{christensen2010dynamo}.  We use the Elasser number rule, which posits that the Lorentz force and Coriolis force are roughly equal, and results in 
\beq
B_\text{core} = \sqrt{\frac{2\,\rho_\text{rock}\,\Omega}{\sigma_\text{electric}}}
\eeq

Here $\Omega$ is the planet's angular rotation speed.

For terrestrial planets, the atmospheric scale height is much smaller than planetary radius, and so it suffices to compare the magnetosphere size to the latter.  Using our expressions above, and defining $\mathcal Y = (R_\text{core}/R_\text{planet})^3$, we find this to be
\beq
\frac{r_\text{magnetosphere}}{R_\text{planet}}=3.1\,\frac{\lambda^{23/60}\,\gamma^{1/6}}{\alpha^{13/12}\,\beta^{11/24}}\,\mathcal{Y}^{1/6}
\eeq

This ratio evaluates to 10 for our values of the constants and Earth's core radius.  The dependence on fundamental constants is rather weak, and so it takes a drastic change to alter the conclusion that the magnetosphere extends beyond the atmosphere.

\subsection{When Is the Temperate Zone Outside the Alfv\'en Zone?}
If a planet is orbiting inside its host star's Alfv\'en zone, its intrinsic magnetic field lines will connect to the star's, which will result in a highly increased level of bombardment by charged particles.  This is expected to be the case for the inner planets in the Trappist-1 system, for instance \citep{garraffo2017threatening} from simulations.  Given our expressions for both the temperate zone and Alfv\'en radius from Section \ref{Properties}, it is straightforward to derive their ratio:
\beq
\frac{a_\text{temp}}{R_A}=7.1\times10^{-4}\,\frac{\lambda^{3/8}\,\gamma^{1/2}}{f_\text{open}\,\alpha^{11/4}\,\beta^2}
\eeq 

For fixed physical constants, this defines a smallest stellar mass for which this condition holds.  In our universe this is about .1 $M_\Sun$, in accordance with the expectation that Proxima Centauri b, which orbits a 0.12 $M_\Sun$ star at 0.05 AU, is outside the Alfv\'en zone for the most likely values inferred for its orbital parameters \citep{garraffo2016space}.  Our treatment ignores the nonsphericity and nonstationarity of the Alfv\'en zone and potential planetary eccentricity, which may cause the orbit to periodically dip into the Alfv\'en zone throughout the year.

Note that the dependence on starspot fraction is of crucial importance in this expression, as otherwise this threshold stellar mass would be smaller than the smallest stellar mass.  As such, this condition is loosely coincident with the onset of a full stellar \mbox{convection zone.}

\subsection{When Does a Core Stratify Geochemically?}
In \citep{unterborn2014role}, it was pointed out that if a planet's mantle oxygen content is too high, the iron will all be in the form of iron oxide (FeO), and will not differentiate to form a core.  They find that the quantity $R_1$ = (Mg + 2Si + Fe)/O must exceed 1 in order for a core to develop, Earth's value of this ratio being 1.23.  Interestingly, about $4\%$ of the Earth's oxygen is left over after binding with magnesium and silicon, so that only $86\%$ of Earth's iron makes it into the core.  This raises the additional possibility that if a planet's oxygen is depleted before its magnesium and silicon are consumed, no iron will be left in the mantle or crust.  This could have an additional adverse effect on habitability, which would restrict the allowable oxygen content required for habitability to a narrow range, but we leave exploration of this for future work.  It was argued in \citep{dyck2021effect} that planets with core mass fraction below $\sim$0.24 would have much higher rates of volatile subduction, due to more extensive volcanism, thicker crust, and stabilized amphibole group.  This places a potential lower limit to the allowable core mass for habitability.

Though the core development condition depends on the ratio $R_1$ above, this depends on the abundances of both the alpha elements and iron, which are set by two different supernova processes, and so will scale differently with fundamental constants. In \citep{mc5}, we found the dependence of the alpha element abundances (C, O, Mg, and Si) on the Hoyle resonance energy $E_R=.626(m_u+m_d)+(0.58\alpha-0.0042)m_p$, with $m_u$, $m_d$ the masses of the up and down quarks, as found in \citep{huang2019sensitivity}.  We also found an expression for the metal to rock ratio, from which we may determine the quantity
\beq
R_2=\frac{\text{Fe}}{\text{Mg+Si+O}}=5.0\times10^{-3}\,\frac{\beta^{.82}\,\gamma^{.54}}{\kappa^{.81}\,\alpha^{.56}}
\eeq 

For Earth, this value is $R_2=0.163$.  This assumes a linear relationship between stellar and planetary metal to rock ratios, which indeed is found \citep{adibekyan2021chemical}.

\begin{figure}[H]
		\includegraphics[width=.7\textwidth]{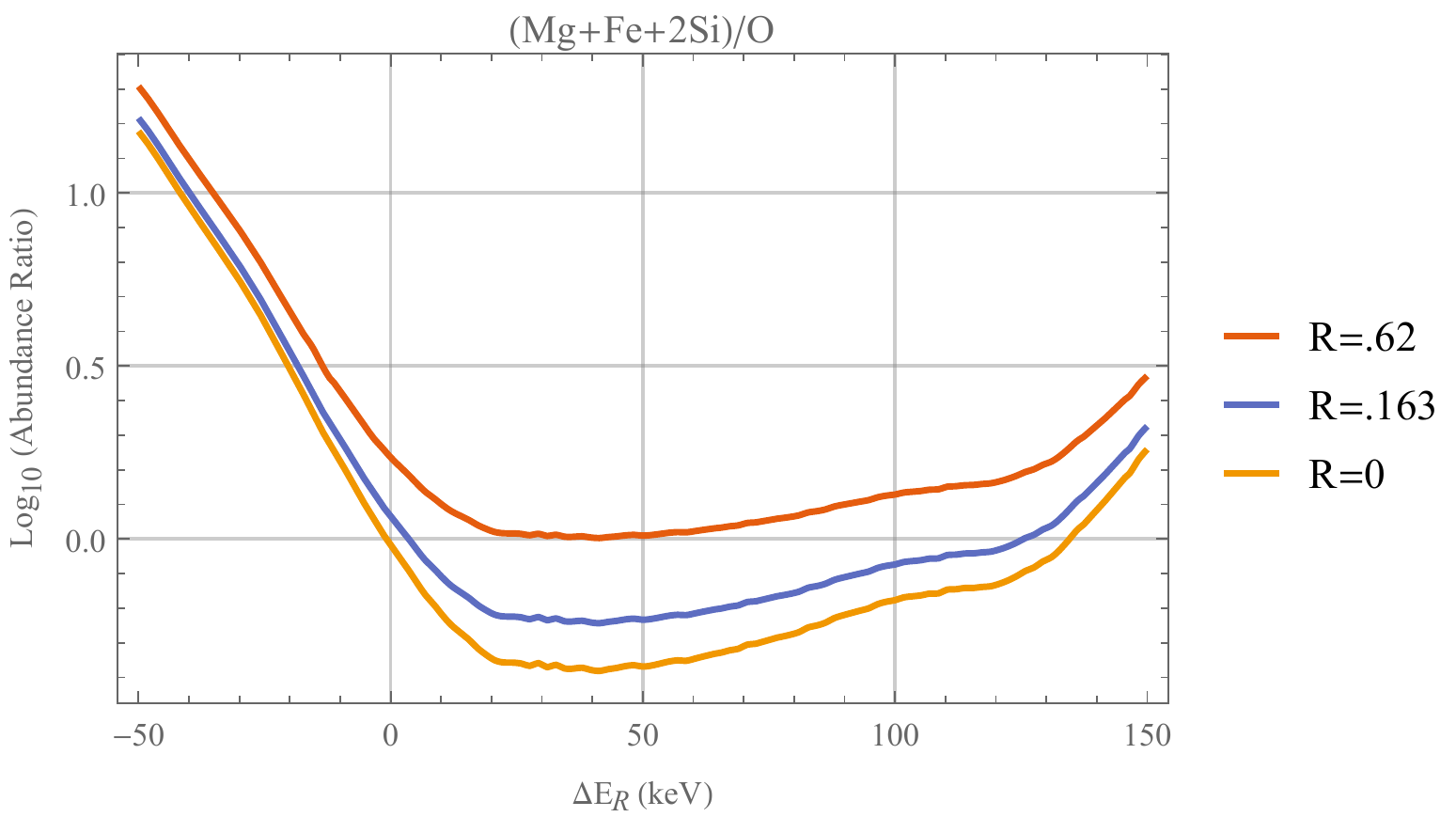}
		\caption{Oxygenation ratio $R_1$ for different metal ratios $R_2$. The quantity $\Delta E_R=0$ for our values of the constants.  Planetary cores form only when this quantity exceeds 1 (0 on our log scale).}
		\label{coreform}
	\end{figure}

In Figure \ref{coreform}, we plot the oxygenation ratio for various values of the metal ratio $R_2$.  It can be seen that with $R_2$ held fixed at the observed value, the oxygenation ratio is less than 1 for $\Delta E_R>3.6$ keV.  While we are rather close to a potential anthropic boundary with metal fraction held fixed, allowing it to vary relaxes this closeness.  In fact, there is a silicon and magnesium rich region of parameter space for larger values of $\Delta E_R$ which also satisfy the $R_1>1$ requirement.  Above a metal fraction of 0.62, these two branches merge, and planets will always contain enough iron to form a core.  As discussed in \citep{mc5}, such metal rich planets may be unsuitable for life for reasons other than the possession of a magnetic field, but we found that placing an upper bound on the metal content does not appreciably affect the probabilities we compute, and we do not concern ourselves with such a boundary here.

\subsection{What Sets the Core Solidification Timescale?}
The presence of a planetary dynamo requires a liquid convective core, which cannot be sustained indefinitely.  As heat leaks from the planet, an initially liquid core will cool and solidify.  If the solidification timescale is too rapid, any magnetic field will cease before life can take hold on a planet, and so one important consideration is the longevity of a \mbox{liquid core.}

First, we must establish that terrestrial planets possess enough heat for their cores to initially be liquid.  This follows almost from our definition of a terrestrial planet, which demands that the gravitational binding energy is of the same order of magnitude as molecular binding energies, so that chemical reactions may take place on the planet's surface.  Given the increased temperature and pressure of the planetary interior, the melting point will naturally be exceeded in the core.

The solidification timescale can be simply estimated as $t_\text{solid}\sim E_\text{core}/Q_\text{heat}$, where $E_\text{core}$ is the energy required to be leached from the core for solidification to take place, and $Q_\text{heat}$ is the total core power.  A proper estimate of $E_\text{core}$ would take into account the difference between the gravitational binding energy and the energy that would result in solidification; thankfully, however, these two energies are similar in magnitude, another consequence of restricting our attention to terrestrial planets.  So, we may approximate the total energy in the core as $E_\text{core}\sim G M_\text{core}^2/R_\text{core}$.  By the same token, $Q_\text{heat}$ has components due to formation and crystallization, which are roughly equal.  In \citep{mc3}, we found that $Q_\text{heat}\sim G M_\text{planet}\rho_\text{rock}\kappa_\text{heat}$ based on dimensional analysis.  There, we also consider radiogenic heat and time dependence in more detail, which we neglect here.  This may indeed be important; as discussed in \citep{nimmo2020radiogenic}, too much radioactive heating can prevent core convection.  However, we do not consider this in detail here.

With this, the core solidification timescale is very simple:
\beq
t_\text{solid}\sim \frac{A_\text{planet}}{\kappa_\text{heat}}=5.7\times10^{-3}\,\frac{M_{pl}^2}{\alpha\,m_e^{5/4}\,m_p^{7/4}}
\eeq

If a long-lived liquid outer core is necessary for habitability, this timescale must be larger than some timescale typical for the development of complex life, which we take here to be proportional to the stellar lifetime (see \citep{mc1} for an exploration of different choices on this matter).  We normalize this time to the expectation that the outer core will remain liquid for another 700 Myr from \citep{airapetian2020impact}.

An alternative view is that a solid inner core is actually necessary for the sustenance of a magnetic field, in spite of geologic evidence to the contrary (see \citep{tarduno2015hadean} for zircon evidence of a magnetic field at 3--4 Ga).  The inner core may have developed as late as 565 Mya, based on magnetic evidence from Ediacaran rocks that record an anomalously low field strength, taken to signal a rearrangement in field configuration indicative of a recently established solid inner core \citep{bono2019young}.  This apparent incompatibility is reconciled if another mechanism generated the magnetic field before core solidification, as for example a long lived liquid mantle ocean \citep{ziegler2013implications}).  In this case, the above timescale would need to be comparable to the evolutionary timescale, rather than simply longer than it.  

\subsection{Is a Planetary Magnetic Field Necessary for Habitability?}
To treat intrinsic planetary magnetic fields as essential for habitability, we include the product of all five factors into the habitability condition as
\bea
\mathbb H_\text{B} = \theta\left(R_a-100\right)\,\theta\left(r_\text{B}-R_\text{planet}\right)\,\theta\left(a_\text{temp}-R_\text{Alfv\'en}\right)
\,\theta\left(R_1-1\right)\,\theta\left(t_\text{solid}-t_\star\right)
\eea

If we incorporate this into our calculation, we find that the Bayes factor relative to the base case where magnetic fields are not taken to be important is $\mathcal B=1.52$.  We also probe the relative importance of each of these subconditions in Table \ref{Btable} by first only incorporating each condition in isolation, and then incorporating the four others without each condition, into the calculation.  Of the five factors considered, the magnetic Reynolds number, magnetosphere radius, and Alfv\'en zone conditions do not perceptibly alter the probabilities.  The core existence condition slightly decreases the probabilities, while the core timescale condition slightly increases them.  So, the notion that a magnetic field is necessary for habitability is compatible with the multiverse, and although it is even slightly preferred to the base case, the difference is not statistically meaningful enough to draw the conclusion that the converse hypothesis is disfavored.

\begin{table}[h]
	\caption{Ablation study for planetary magnetic field criteria.  This table displays the Bayes factors relative to the baseline case where planetary magnetic fields are not important.  The `with only' rows only incorporate the condition in the given column, and the `without only' rows incorporate every condition {\it except} the condition in  the given column into the probability calculation.}
	\label{Btable}
	\setlength{\cellWidtha}{\textwidth/6-2\tabcolsep+0.0in}
\setlength{\cellWidthb}{\textwidth/6-2\tabcolsep+0.0in}
\setlength{\cellWidthc}{\textwidth/6-2\tabcolsep+0.0in}
\setlength{\cellWidthd}{\textwidth/6-2\tabcolsep+0.0in}
\setlength{\cellWidthe}{\textwidth/6-2\tabcolsep+0.0in}
\setlength{\cellWidthF}{\textwidth/6-2\tabcolsep+0.0in}
\scalebox{1}[1]{\begin{tabularx}{\textwidth}{>{\PreserveBackslash\centering}m{\cellWidtha}>{\PreserveBackslash\centering}m{\cellWidthb}>{\PreserveBackslash\centering}m{\cellWidthc}>{\PreserveBackslash\centering}m{\cellWidthd}>{\PreserveBackslash\centering}m{\cellWidthe}>{\PreserveBackslash\centering}m{\cellWidthF}}
			\toprule
			\pmb{$\mathbb H$} & \boldmath{$R_a$} & \boldmath{$r_B>R_\text{\textbf{planet}}$} & \boldmath{$a_\text{\textbf{temp}}>R_A$} & \textbf{Mg + 2Si + Fe > O} & \boldmath{$t_\text{\textbf{solid}}>t_\star$} \\
			\midrule
			\multicolumn{6}{l}{Earth-like C/O}\\
			\midrule
			with only & 1.0 & 1.0 & 1.0 & 0.658 & 1.19\\
			without only & 1.52 & 1.52 & 1.44 & 1.25 & 0.68\\
			\midrule
			\multicolumn{6}{l}{Unrestricted C/O}\\
			\midrule
			with only & 0.24 & 1.98 & 1.99 & 2.12 & 0.0046\\
			without only & 0.050 & 0.050 & 0.046 & 0.0043 & 1.73\\
			\bottomrule
		\end{tabularx}}
\end{table}

We also remark that the base case here took the carbon-to-oxygen ratio to be important for habitability.  If instead we drop this assumption, we find the Bayes factor is $\mathcal B = 0.050$.  The driving factor in making this so low is the core solidification timescale, as can be seen in Table \ref{Btable}.  Therefore, we find that the assumption that planetary magnetic fields are important is only compatible with the multiverse if carbon-to-oxygen ratio is also important.  This echoes the results of Section \ref{Atmosphere} and \citep{mc5}, where we found that restricting the carbon-to-oxygen ratio was important for compatibility with the multiverse on other accounts.  This also suggests a test of the multiverse hypothesis, for if we find that complex life occurs only on magnetized planets but independently of carbon-to-oxygen ratio, our presence in this universe would be quite unlikely.

\section{Discussion}

Though few agree on exactly what conditions are required for habitability, it surely depends on the confluence of a great many factors.  Likewise, our notions of habitability strongly affect the expectation for the distribution of life, both throughout our universe, and in others.  Because of this, very fine-grained effects have the potential to radically alter our estimations of the probability of our existence in this particular universe, and our observations in general.  This places us in a scenario where the importance of all discussed habitability factors must be tested before we may make any statements about multiverse probabilities with a relatively high degree of certainty.  The stellar activity and atmospheric aspect of this program was undertaken in this paper.

Uncertainties abound: the physics dictating the corona, stellar wind and flares, the relative importance of different atmospheric erosion rates, the ultimate source of Earth's atmosphere, the importance of planetary magnetic fields, and the distribution of all these quantities across different stellar systems are only now coming to light.  While we have tried to hedge our ignorance in as many aspects as possible by contemplating competing accounts of these effects, we have necessarily restricted our attention in certain cases, and completely neglected other potentially important effects.  Thus, while our work cannot claim to be a definitive exploration of stellar activity and atmospheric effects in other universes, it does represent an important first step.

Perhaps the biggest takeaway of our findings is that, if one believes that a relatively narrow carbon-to-oxygen ratio is required for complex life (as may be argued by the vastly different tectonic regimes that occur outside the interval (0.5,1)), any atmospheric habitability condition we considered had no significant bearing on multiverse probabilities.  In this light, all that can be said is that atmospheric presence and stability does not appear to be a major determining factor for why we are in this universe.  This is plausible, since the Earth's atmosphere is about two orders of magnitude larger than any threshold we are aware of, but many effects we consider depend exponentially on fundamental constants, so this conclusion is by no means automatic.

On the other hand, if we entertain the possibility that a carbon-to-oxygen ratio relatively close to ours is not required for habitability, altogether different conclusions are drawn.  We are forced to conclude, under this assumption, that planetary magnetic fields cannot be important for life, because it renders many of the otherwise less likely regions of parameter space infertile, making us outliers.  Additionally, when treating the carbon-to-oxygen ratio as unimportant, we find planetary atmospheric nitrogen must not scale with stellar system nitrogen abundance, or our presence in this universe is unlikely, independent of uncertainties about atmospheric source and lower atmospheric mass threshold.

Both of these findings also suggest potential methods for testing the multiverse hypothesis, if the true habitability conditions turn out to be incompatible with these expectations.  So, if we find that an Earth-like C/O is not needed for complex life and either that atmosphere mass scales with stellar nitrogen or that planetary magnetic fields are required for life, the predictions the multiverse framework has made will be found to be incorrect.  While these tests may be rather far off, the salient point is that they are possible in principle.  {Various biosignatures have already been proposed to help determine the distribution of life throughout the universe, for several places inside and out of our solar system, including searching for relic biomarker compounds on Mars} \citep{westall2015biosignatures}, {abundance ratios of organic compounds on icy moons such as Enceladus} \citep{mckay2008possible,taubner2018biological}, {chemical disequilibrium and even microbial absorption in Venus's atmosphere} \citep{limaye2021venus}, {and atmospheric gases such as oxygen around exoplanets} \citep{meadows2017reflections}.  In fact, it is conceivable that the next few generations of experiments will be able to measure biosignatures on exoplanet populations large enough to distinguish trends with respect to system parameters such as composition \citep{schwieterman2018exoplanet}, that the presence of planetary magnetic fields can be measured through auroral emissions \citep{green2021magnetospheres}, and that the relation between planetary atmospheric size and stellar composition can be determined \citep{crossfield2017trends}.

\vspace{6pt}
\authorcontributions{Conceptualization, all authors; 
Methodology, M.S.; 
Formal Analysis, M.S.; 
Validation, V.A., L.B. and G.F.L.; 
Writing---Original Draft Preparation, M.S.; 
Writing---Review $\&$ Editing, V.A., L.B. and G.F.L. All authors have read and agreed to the published version of the manuscript.}

\funding{This research received no external funding.}

\dataavailability{All code to generate data and analysis is located at \url{https://github.com/mccsandora/Multiverse-Habitability-Handler}, accessed on Dec. 20, 2022.} 

\acknowledgments{We would like to thank Daman Grewal for useful comments.}

\conflictsofinterest{The authors declare no conflict of interest.} 


\begin{adjustwidth}{-\extralength}{0cm}

\reftitle{References}

\PublishersNote{}
\end{adjustwidth}
\end{document}